\documentclass{PoS}

\title{Deflated Hermitian Lanczos Methods for Multiple Right-Hand Sides}

\ShortTitle{Deflated Hermitian Lanczos Methods}

\author{Abdou M. Abdel-Rehim\thanks{New address: 
Department of Physics, The College of William \& Mary} \\
Department of Physics, Baylor University, Waco, TX 76798-7316, USA \\
E-mail: \email{amrehim@cs.wm.edu}}
\author{Ronald B. Morgan \\
Department of Mathematics, Baylor University, Waco, TX 76798-7328, USA \\
E-mail: \email{ronald\_morgan@baylor.edu}}
\author{Dywayne Nicely \\
Department of Mathematics, Baylor University, Waco, TX 76798-7328, USA \\
E-mail: \email{dywayne\_nicely@baylor.edu} }
\author{\speaker{Walter Wilcox} \\
Department of Physics, Baylor University, Waco, TX 76798-7316, USA \\
E-mail: \email{walter\_wilcox@baylor.edu}}

\abstract{A deflated and restarted Lanczos algorithm to solve hermitian linear systems, and at the same time compute eigenvalues and eigenvectors for application to multiple right-hand sides, is described. For the first right-hand side, eigenvectors with small eigenvalues are computed while simultaneously solving the linear system. Two versions of this algorithm are given. The first is called Lan-DR and is based on conjugate gradient (CG) implementation of the Lanczos algorithm. This version will be optimal for the hermitian positive definite case. The second version is called MinRes-DR and is based on the minimum residual (MinRes) implementation of Lanczos algorithm. This version is optimal for indefinite hermitian systems where the CG algorithm is subject to instabilities. For additional right-hand sides, we project over the calculated eigenvectors to speed up convergence. The algorithms used for subsequent right-hand sides are called D-CG and D-MinRes respectively. After some introductory examples are given, we show tests for the case of Wilson fermions at kappa critical. A considerable speed up in the convergence is observed compared to unmodified CG and MinRes.}

\FullConference{The XXVI International Symposium on Lattice Field Theory\\
		 July 14-19 2008\\
		 Williamsburg, Virginia, USA}

\begin{document}

\section{CG and MinRes Examples}

We have presented and described deflation methods for non-hermitian systems in previous talks\cite{1} and papers\cite{2}. These methods simultaneously solve the linear equations and compute eigenvalues and eigenvectors. Recently, we have also developed similar methods for hermitian systems\cite{3}, and we will partially describe that work here. See also Ref.~\cite{4} for a description of a new approach for solving multiple right-hand sides using hermitian \lq\lq seed" methods.

Krylov subspace methods develop polynomial solutions of $Ax=b$ for a hermitian matrix $A$ given an initial residual vector $r_0$. This polynomial space, $K$, after $m$ iterations of a method like conjugate gradient (CG) or minimum residual (MinRes), is given by
\begin{equation}
K=Span\{r_0, Ar_0,  A^2r_0,... A^{m-1}r_0\}.
\end{equation}
The polynomial produced, considered as a continuous function in eigenvalue space, $\lambda$, is of degree $m$ or less and has the value 1 at $\lambda=0$. If the problem is solved exactly, one can show that this polynomial has a zero at the position of each eigenvalue, $\lambda_i$. To get a feeling for the types of behavior expected from the standard CG and MinRes routines, consider a hermitian problem of dimension 1000 with positive definite eigenvalues $\lambda_i=.1,1,2,3,\dots , 999$. The CG and MinRes polynomials developed after 70 iterations are shown in Fig.~1. The eigenvalues are circled in blue. The CG polyniomial is closer to zero at the lowest eigenvalue, but the MinRes polynomial is better at zeroing out most of the smaller eigenvalues. The residual norm curves for these solutions, as a function of iteration number, are shown in Fig.~2. Notice that the better MinRes eigenvalue polynomial is reflected in the slightly faster convergence versus CG.

\begin{figure}
\begin{center}
\leavevmode
\includegraphics*[scale=0.5]{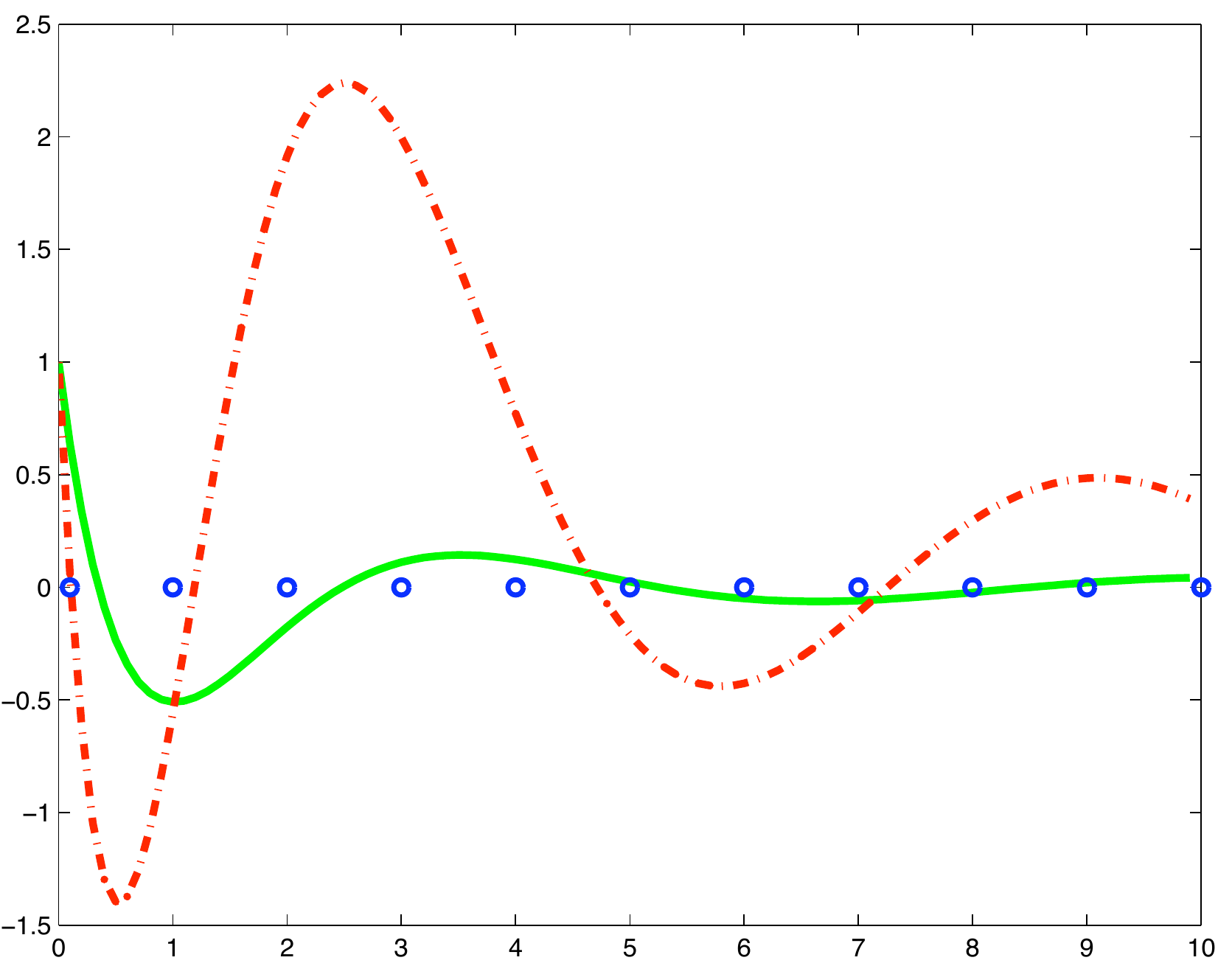}
\caption{MinRes (solid green) and CG (dot dashed red) polynomials of degree 70 in eigenvalue space on a small problem with a positive definite spectrum of dimension 1000. The real eigenvalues are shown with blue circles.}
\vspace{1cm}
\includegraphics*[scale=0.5]{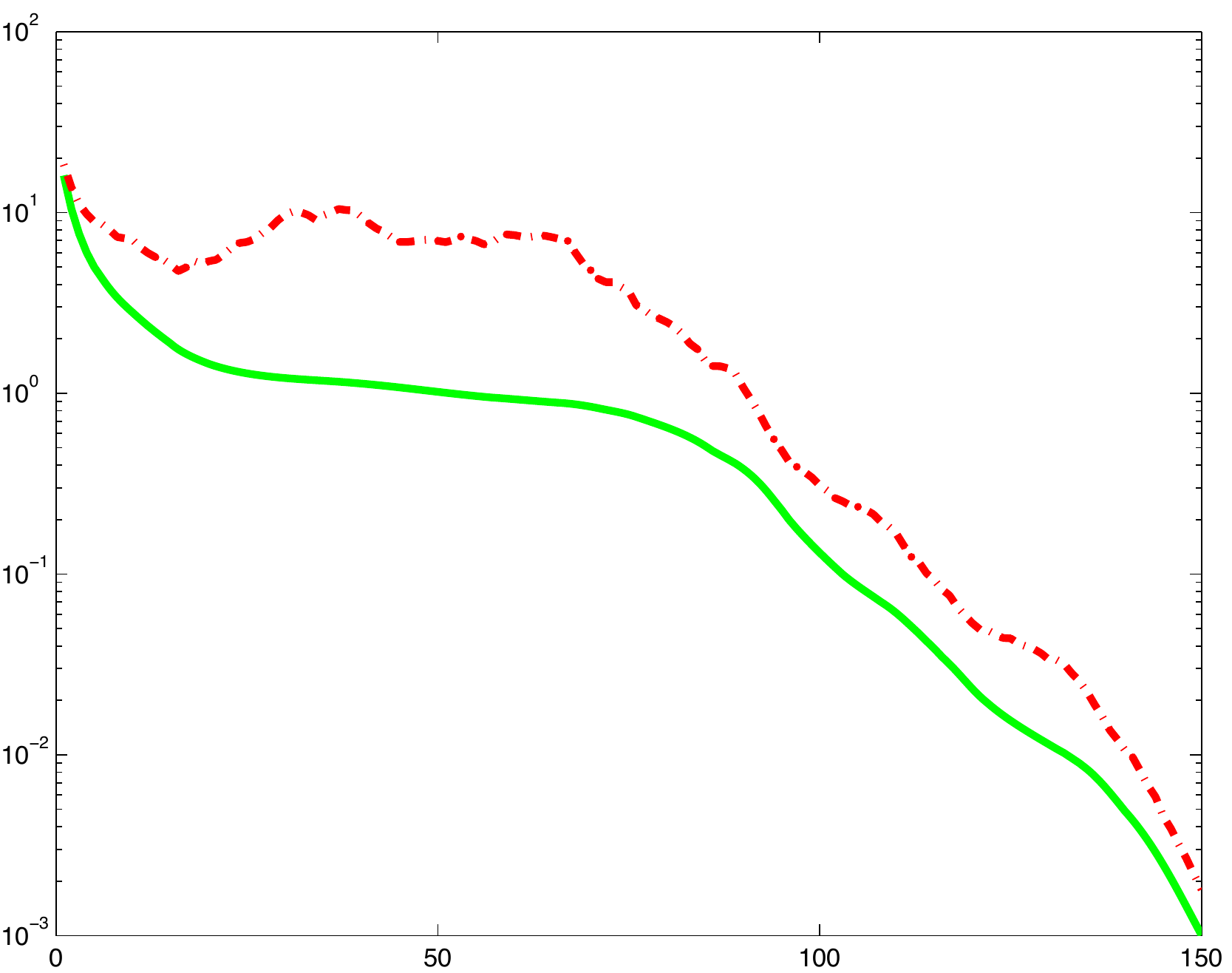}
\caption{Residual norm curves for MinRes (solid green) and CG (dot dashed red) on the problem in Fig.~1.}
\end{center}
\end{figure}

Let us consider another example, which has an indefinite eigenvalue spectrum and shows the effects of deflation. Fig.~3 shows the MinRes and CG polynomials developed after 110 iterations for a hermitian system of dimension 1000 whose diagonal entries are generated with random numbers distributed Normal (0,1), but shifted by 2.0 to the right. There are then 22 negatives among the 1000 eigenvalues. Both algorithms have polynomials with value 1 at $\lambda=0$. One can again see the MinRes polynomial is doing a better job of zeroing the eigenvalue spectrum than the CG one. Fig.~4 shows the resultant residual norm curves. MinRes slightly outperforms CG again. This indefinite spectrum problem is more difficult for CG, resulting in a spiked residual norm curve. Notice that both methods display so-called super-linear convergence after about 125 iterations, when the small eigenvalues in the spectrum are effectively removed or deflated out. This is also referred to as the deflation \lq\lq knee". A method which retained and used such deflated eigenvector information to solve additional right-hand sides, $b$, of $Ax=b$, clearly would be greatly speeded up.

Both of these standard methods are, of course, unrestarted. For similar restarted methods, it will be important to keep the low eigenvalue information across the restart so that the Krylov subspace polynomials will not have to redevelop it. Thus, to effectively implement deflation of small eigenvalues in restarted methods, it is indispensable to simultaneously solve the small eigenvalue/eigenvector problem along with the linear equations. As a bonus, this low eigenvalue information can then be used to speed up the solution of subsequent right-hand sides, $b$, of $Ax=b$. We implement these ideas in a hermitian context with a deflated Lanczos/conjugate gradient algorithm combination, Lan-DR($m,k$)/D-CG, which is optimal for a positive definite spectrum. Similarly, a deflated minimum residual combination, MinRes-DR($m,k$)/D-MinRes, is designed to be optimal for indefinite hermitian systems. Here $m$ stands for the dimension of the Krylov subspace and $k$ is the number of deflated eigenvectors. The second algorithm in the two cases, either D-CG or D-MinRes, refers to the form of the algorithm that projects over these deflated eigenvectors to speed up the solution for additional multiple right-hand sides. We will see that the general behaviors for CG or MinRes in the simple examples considered will carry over to the new algorithms on lattice QCD problems. Please see Ref.~\cite{3} for the mathematical definition of these algorithms and more details.

\begin{figure}
\begin{center}
\leavevmode
\includegraphics*[scale=0.5]{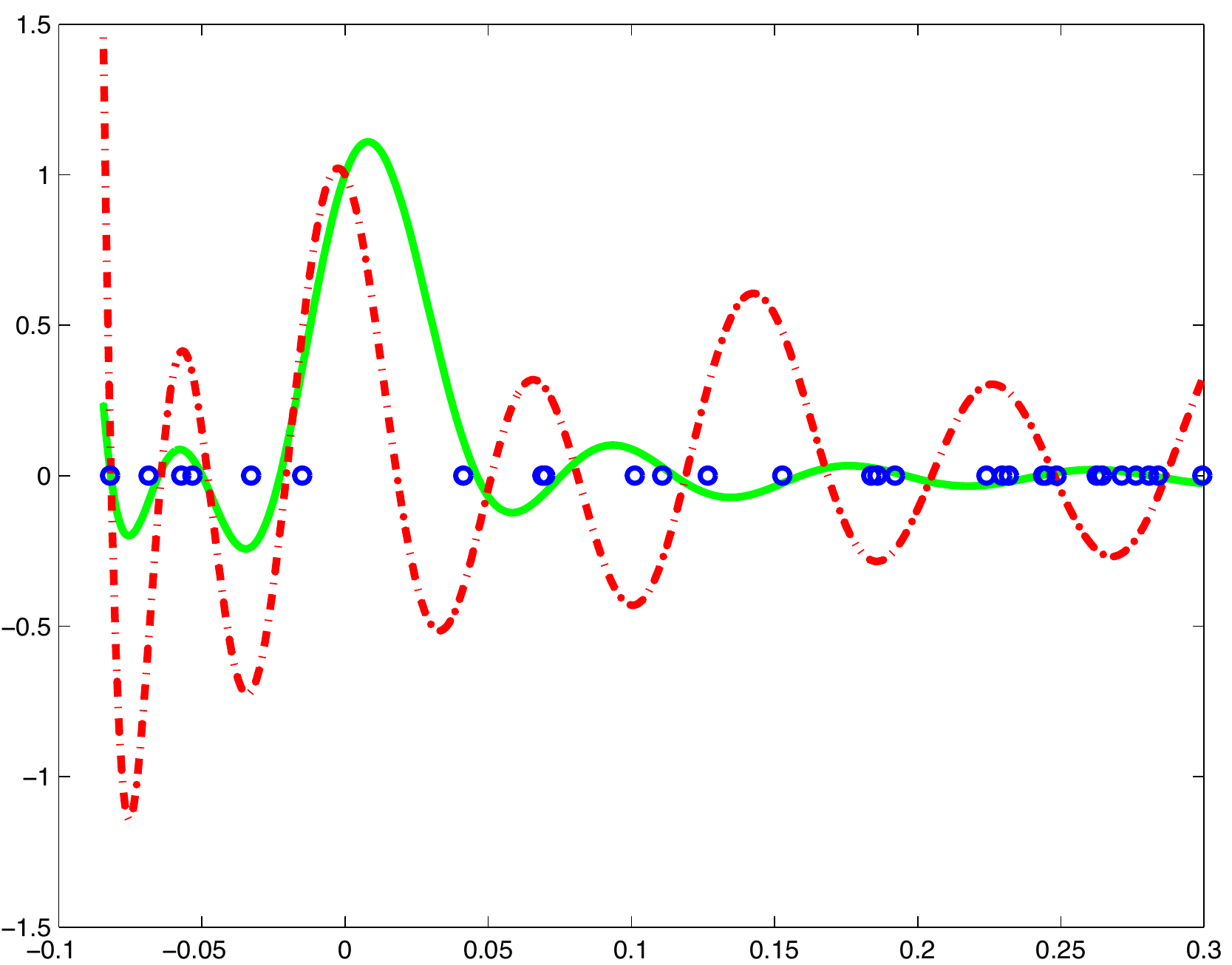}
\caption{MinRes (solid green) and CG (dot-dashed red) polynomials of degree 110 in eigenvalue space on a small indefinite spectrum problem with dimension 1000. The real eigenvalues are shown with blue circles.}
\end{center}
\end{figure}

\begin{figure}
\begin{center}
\leavevmode
\includegraphics*[scale=0.5]{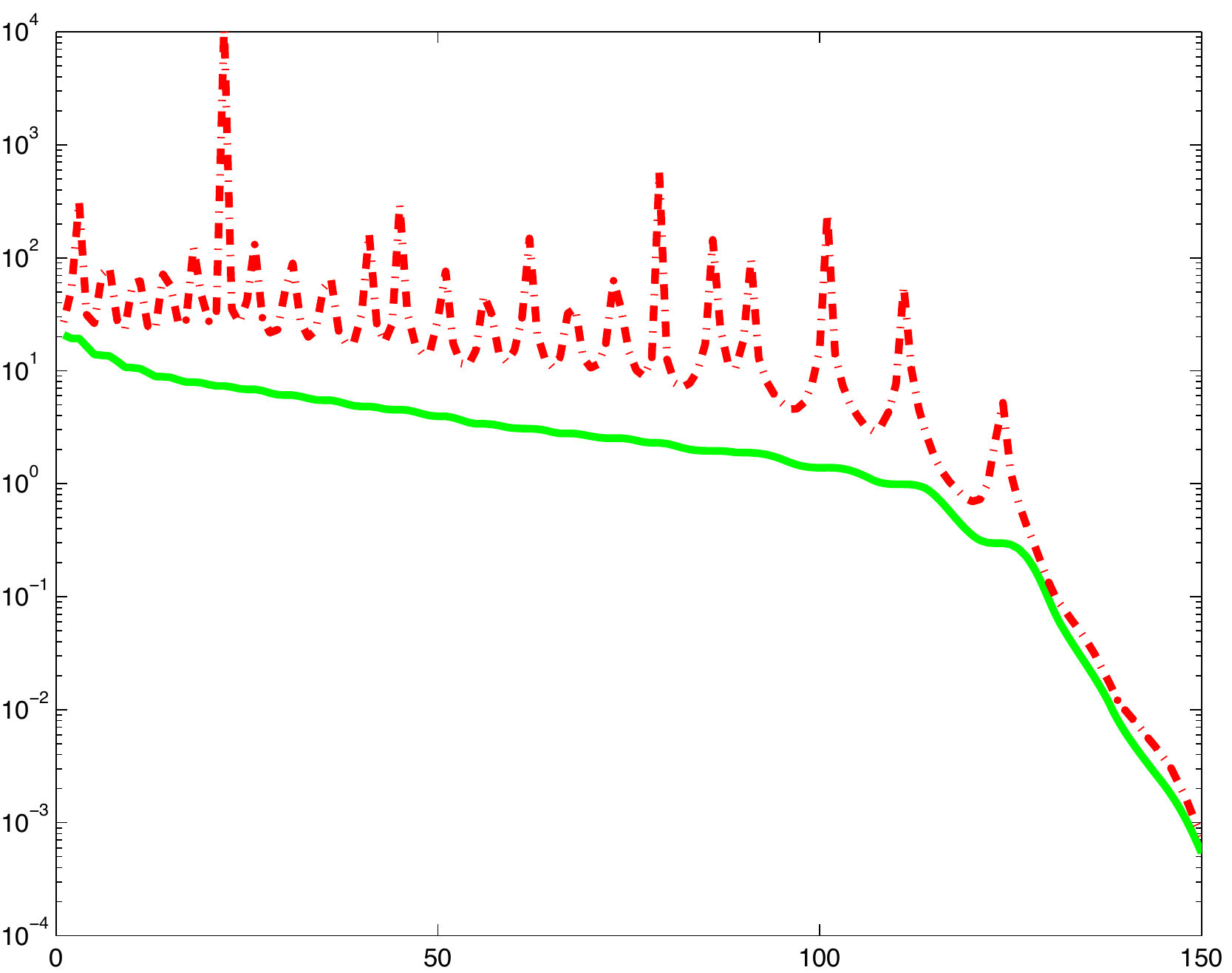}
\caption{Residual norm curves for MinRes (solid green) and CG (dot-dashed red) on the problem in Fig.~3.}
\end{center}
\end{figure}

\begin{figure}
\begin{center}
\leavevmode
\includegraphics*[viewport=0 0 800 530, scale=0.4]{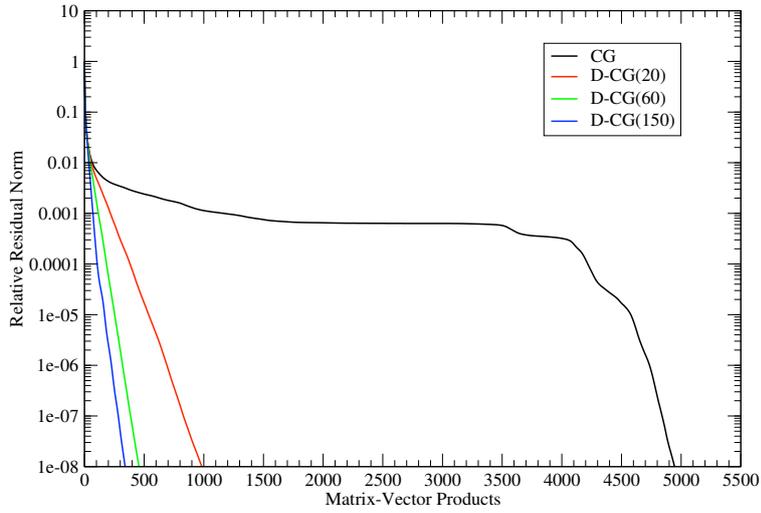}
\caption{D-CG convergence curves compared to CG on a $\beta=6.0$ quenched $20^3\times 32$ lattice at $\kappa=0.15720$, solving through $M^{\dagger}Mx=M^{\dagger}b$.}
\label{D-CG}
\end{center}
\end{figure}

\begin{figure}
\begin{center}
\leavevmode
\includegraphics*[viewport=0 0 800 530, scale=0.4]{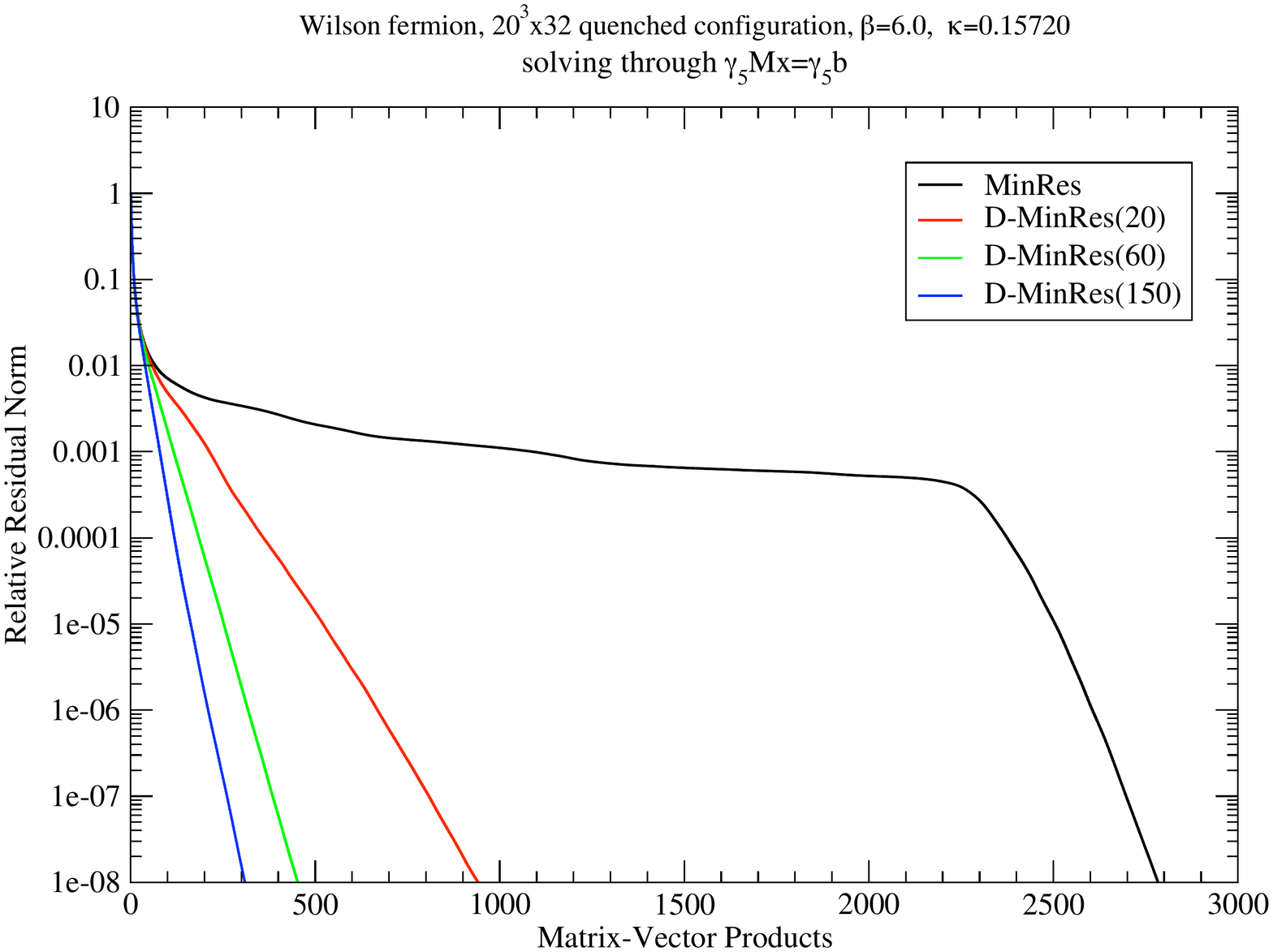}
\caption{D-MinRes convergence curves compared to MinRes on the same configuration and quark mass as Fig.~5, solving through $\gamma_5 Mx=\gamma_5 b$.}
\end{center}
\end{figure}

\section{Lattice Applications}

In lattice applications, as in the simple examples considered above, we would expect the Lan-DR($m,k$)/D-CG combination to be effective mainly on problems with positive definite spectra, while the MinRes-DR($m,k$)/D-MinRes combination is designed for indefinite spectra. Designating $M$ as the Wilson matrix, we will consider $M^{\dagger}M$ as a model for the former spectrum and $\gamma_5 M$ as a model for the latter in our tests. The following examples show the residual norm for convergence of $M^{-1}$ itself so that results from the various algorithms can be compared. The residual norm quoted is normalized to one for an initial guess of $x=0$ for the solution vector. These runs are on quenched $\beta=6.0$ $20^3\times 32$ lattices at essentially kappa critical ($\kappa =0.15720$).

Fig.~5 shows the residual norm as a function of matrix-vector products (MVPs) of the $M^{\dagger}M$ problem using CG. Although the convergence curves are not shown, the Lan-DR($m,k$) results, done for $m=200$ and various $k$ values (20, 60, or 150), are quite similar to the CG result, which took about 4900 MVPs to converge to a relative residual norm of $10^{-8}$. (The actual number of MVPs for Lan-DR to reach the same level of convergence ranged from about 5300 for $k=20$ to about 4950 for $k=150$.) The eigenvectors are identified from the solution of the exterior eigenvalue problem using regular Ritz vectors and are then passed on to D-CG, which uses a Galerkin projection to deflate out these eigenvectors. The next right-hand side is then greatly accelerated. Notice the sharp deflation knee occurring at $\sim 4000$ iterations and the subsequent super-linear convergence. The slope of this curve after the knee is the rate full deflation will achieve when a sufficient number of eigenvectors are kept. We find that full convergence to $10^{-8}$ is achieved in a little over 300 iterations when 150 eigenvectors are deflated on additional right-hand sides. We would not expect to see additional acceleration from deflating more eigenvectors since the rate of convergence approximately matches the slope of the CG curve after the deflation knee. We have referred to this phenomenon as \lq\lq saturation" (first ref. in \cite{1}).

\begin{figure}
\begin{center}
\leavevmode
\includegraphics*[scale=0.5]{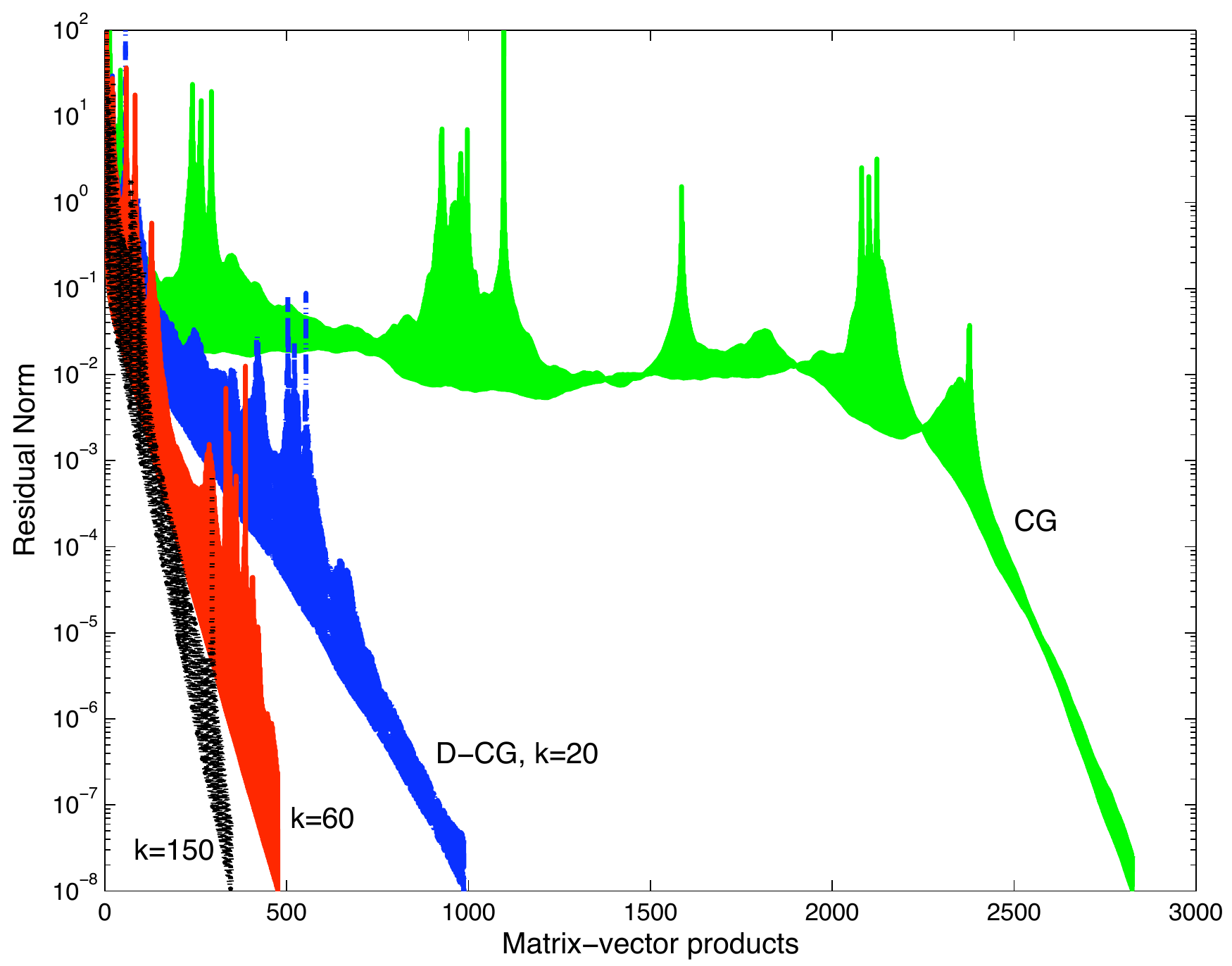}
\caption{D-CG convergence curves compared to CG on the same configuration and $\kappa$ as Fig.~5, solving through $\gamma_5 Mx=\gamma_5 b$.}
\end{center}
\end{figure}

Fig.~6 shows the residual norm of the MinRes algorithms on the same lattice matrix. Again, the MinRes($m,k$) results, with $m=200$ and $k=20, 60$ or $150$ are not shown but converge similarly to the unrestarted MinRes result, which took about 2800 MVPs to converge to a relative residual norm of $10^{-8}$. (The actual number of MVPs for MinRes-DR to reach the same level of convergence ranged from about 3100 for $k=20$ to about 2900 for $k=150$.) Note that the unrestarted MinRes solution takes fewer MVPs compared to pure CG since there are two MVPs per iteration of the CG normal equations as opposed to the one with MinRes. The small eigenvectors are identified from the harmonic Ritz vectors, which are passed on to the D-MinRes algorithm. Similar to D-CG, a Galerkin projection is again applied at the beginning of the algorithm for multiple right-hand sides. In this case, we obtain full convergence of the residual norm to $10^{-8}$ in $\sim 300$ iterations with 150 eigenvectors, which is slightly faster than with D-CG.

We do not recommend the use of the Lan-DR/D-CG combination for indefinite spectra. We see the effect of such usage in Fig.~7, which shows the use of CG on the original right-hand side (Lan-DR would be similar) and D-CG on additional right-hand sides of the same matrix used ($\gamma_5 M$) in Fig.~5. The extremely spiked nature of the residual norms is evidence of the instability, similar to the example in Fig.~4. For this particular configuration the residual norm converged. We have also seen cases where it does not converge. However, although the first right-hand side may not converge, we have observed that the $k$ deflated eigenvectors and eigenvalues output still produce accelerated convergence for additional right-hand sides.

To avoid roundoff errors associated with the Lanczos algorithm we apply a periodic re-orth-ogonalization over the eigenvectors kept at restart during the solution of the first right-hand side. Also, please note that to achieve the full effect of the deflated eigenvectors for additional right-hand sides, we typically run the first right-hand side (using Lan-DR or MinRes-DR) well past convergence of the linear equations. This overhead contributes at most a factor of two in MVPs on the first right-hand side, and often less.

\section{Summary}

Lan-DR($m,k$) combines the solution of linear equations with the calculation of $k$ Ritz eigenvectors. It is optimal for postive definite spectra. If calculated to sufficient precision, it provides D-CG an efficient starting point for multiple right-hand sides. Minres-DR$(m,k$)/D-Minres does the same for systems with an indefinite spectrum using harmonic Ritz vectors. Although the Lan-DR/D-CG combination is simplier to implement, both MinRes-DR and D-Minres converged slightly faster than their counterparts as a function of MVPs for our tests with the quenched Wilson matrix.

\section{Acknowledgment}

Calculations were done with the HPC systems at Baylor University.


\begin{thebibliography}{99}
\bibitem{1} W. Wilcox, \pos{PoS(Lattice 2007)025}; A. Abdel-Rehim, R. B. Morgan, and W. Wilcox, \pos{PoS(Lattice 2007)026}; D. Darnell, R. B. Morgan, and W. Wilcox, Nucl. Phys. Proc. Suppl. {\bf 129} (2004) 856; R. B. Morgan and W. Wilcox, Nucl. Phys. Proc. Suppl. {\bf 106} (2002) 1067.
\bibitem{2} R. B. Morgan and W. Wilcox, {\it Deflated Iterative Methods for Linear Equations with Multiple Right-Hand Sides}, submitted to: Elec. Trans. on Num. Anal. (arXiv: 0707.0505); D. Darnell, R. B. Morgan, and W. Wilcox, {\it Deflated GMRES for Systems with Multiple Shifts and Multiple Right-Hand Sides}, Lin. Alg. and its Appl., {\bf 429} (2008) 2415; R. B. Morgan and W. Wilcox, arXiv: math-ph/0405053.
\bibitem{3} A. Abdel-Rehim, R. B. Morgan, D. A. Nicely, and W. Wilcox, {\it Deflated and Restarted Symmetric Lanczos Methods for Eigenvalues and Linear Equations with Multiple Right-Hand Sides}, submitted to: SIAM J. Sci. Comp. (arXiv: 0806.3477).
\bibitem{4} A. Abdel-Rehim, R. B. Morgan, and W. Wilcox, {\it Seed Methods for Linear Equations in Lattice QCD Problems with Multiple Right-Hand Sides}, \pos{PoS(Lattice 2008)038}.

\end{thebibliography}
\end{document}